\documentclass[11pt, oneside]{article}
\pdfoutput=1
 
\usepackage{geometry}
\usepackage{graphicx}
\usepackage{changebar}
\usepackage{amssymb}
\usepackage{enumitem}
\usepackage{hyperref}
\usepackage{scrextend}

\def\figlayout#1{\[\centerline{\hbox{\includegraphics[width=\hsize]{#1}}}\]}
\def\figsizeW#1#2{\[\centerline{\hbox{\includegraphics[width=#2]{#1}}}\]}
\def\figsizeH#1#2{\[\centerline{\hbox{\includegraphics[height=#2]{#1}}}\]}

\def\doiUrl#1{\href{https://doi.org/#1}{{\it doi:} #1}}

\title{On the Unimportance of Superintelligence}
\author{John G. Sotos, MD MS(AI) \\ Colonel, US Air Force (Ret.) \\ August 29, 2021}
\date{}

\begin{document}
\maketitle

\section*{\bf Abstract}

\noindent Humankind faces many existential threats, but has limited resources to mitigate them. Choosing how and when to deploy those resources is, therefore, a fateful decision.  Here, I analyze the priority for allocating resources to mitigate the risk of superintelligences.

Part I observes that a superintelligence unconnected to the outside world (de-efferented) carries no threat, and that any threat from a harmful superintelligence derives from the peripheral systems to which it is connected, e.g., nuclear weapons, biotechnology, etc.  Because existentially-threatening peripheral systems already exist and are controlled by humans, the initial effects of a superintelligence would merely add to the existing human-derived risk. This additive risk can be quantified and, with specific assumptions, is shown to decrease with the square of the number of humans having the capability to collapse civilization.

Part II proposes that biotechnology ranks high in risk among peripheral systems because, according to all indications, many humans already have the technological capability to engineer harmful microbes having pandemic spread.  Progress in biomedicine and computing will proliferate this threat. ``Savant'' software that is not generally superintelligent will underpin much of this progress, thereby becoming the software responsible for the highest and most imminent existential risk -- ahead of hypothetical risk from superintelligences.

The analysis concludes that resources should be preferentially applied to mitigating the risk of peripheral systems and savant software. Concerns about superintelligence are at most secondary, and possibly superfluous.

\section*{Introduction}

Many scholars believe that a highly capable advanced intelligence -- a ``superintelligence'' -- poses a near-term existential threat to human civilization.\footnote{Herein, ``existential threat'' refers to the collapse of civilization or the extinction of humankind [Beard et al]. Although ``collapse'' has an accepted socio-political definition [Tainter], it here means a profound loss of technological capacity, e.g., to a level where interstellar communication is not possible [Sagan].}  This essay demonstrates that these concerns, well-meaning as they may be, are harmful because, first, they distract attention from the true loci of danger and, second, because they underestimate the proximity of the threat from less capable  ``savant'' systems already in existence.

As machine intelligences have attracted the most concern, our discussion will be restricted to such intelligences.  Extrapolation to other substrates of intelligence is straightforward. Though interdisciplinary, this paper aims for an audience oriented more toward computers than biosciences, and thus simplifies many of the biotechnological concepts.

\part{Decomposing Risk}

\section{Superintelligences and Savants}

A superintelligence is ``any intellect that greatly exceeds the cognitive performance of humans in virtually all domains of interest'' [Bostrom, p26] [Good].

By contrast, we may define a ``savant'' as any computing system that greatly exceeds the cognitive performance of humans in only one or a few domains of interest (Figure~1).
Savants may range from low-complexity software to specialized artificial intelligence. For example, a pocket calculator and a protein-folding predictor both qualify as savant software, albeit of greatly differing complexities.\footnote{An alternative terminology calls savants ``narrow superintelligence'' and defines ``general superintelligence'' to be what the present work calls simply ``superintelligence.'' The alternative terminology seems undesirable because it is wordier and corrupts the Bostrom/Good definition of ``superintelligence'' given above.}

\begin{figure}[htbp]
   \centering
   \figsizeH{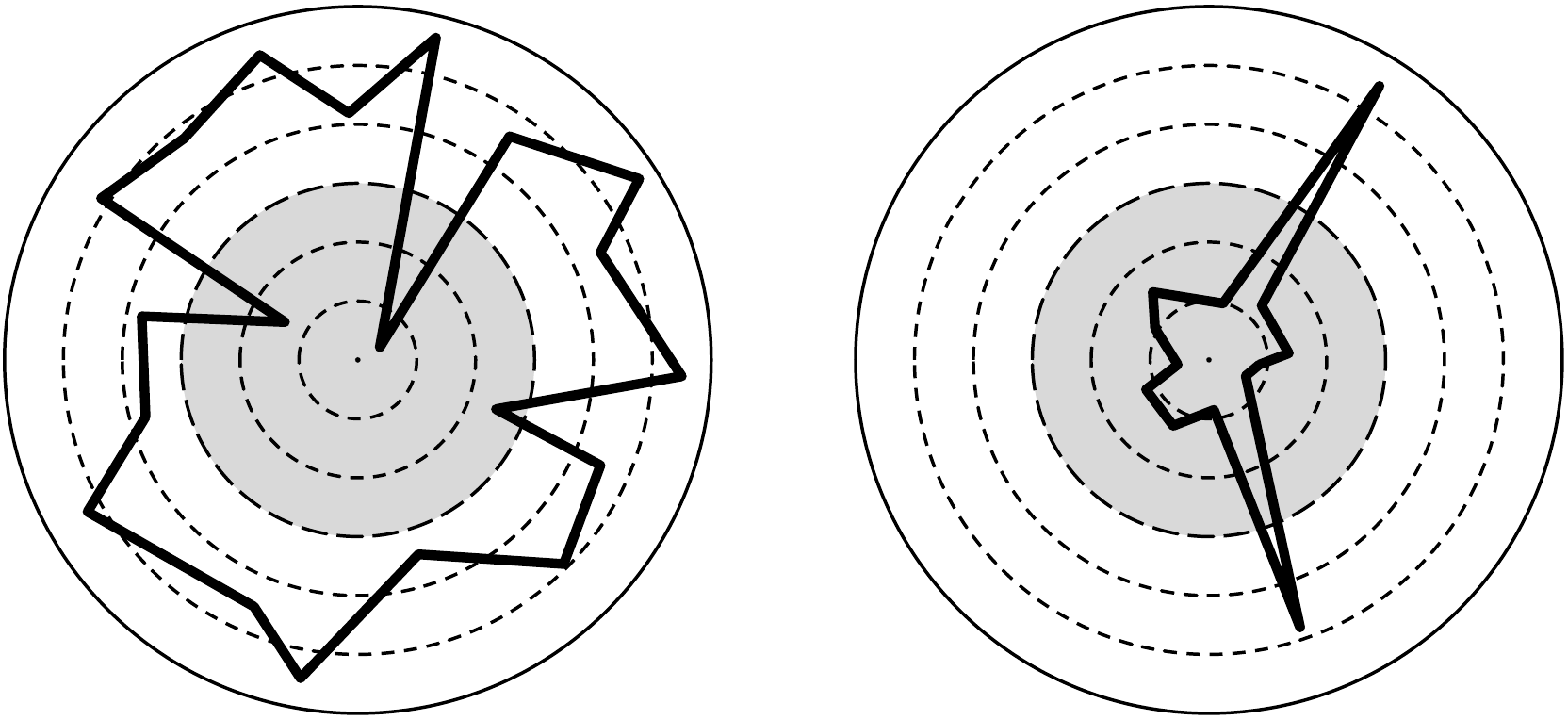}{1in}
   \vskip -16pt
   \caption{\scriptsize Radar diagrams of a superintelligence (left) and a savant (right). Cognitive performance increases with distance from the center. Heavy black lines show each system's cognitive performance for several domains, each vertex being a domain. The gray circle is the cognitive performance of a human for the same domains. In this hypothetical example, the savant exceeds human performance for only two domains, the superintelligence for all but three of the 20 domains plotted.}
   \label{fig:radar}
\end{figure}

\section{Central Systems and Peripheral Systems}

An intelligence, no matter how smart, how dull, how benevolent, or how evil, cannot affect human lives if it is walled off by itself and able neither to communicate with the outside world nor physically manipulate the outside world.\footnote{We ignore power consumption, waste production, and other sustaining functions normally unassociated with intelligence.} The medical profession would apply the descriptors ``locked-in'' or ``de-efferented'' to such isolated intelligences [Plum \&\ Posner].

\begin{figure}[htbp]
   \centering
   \figlayout{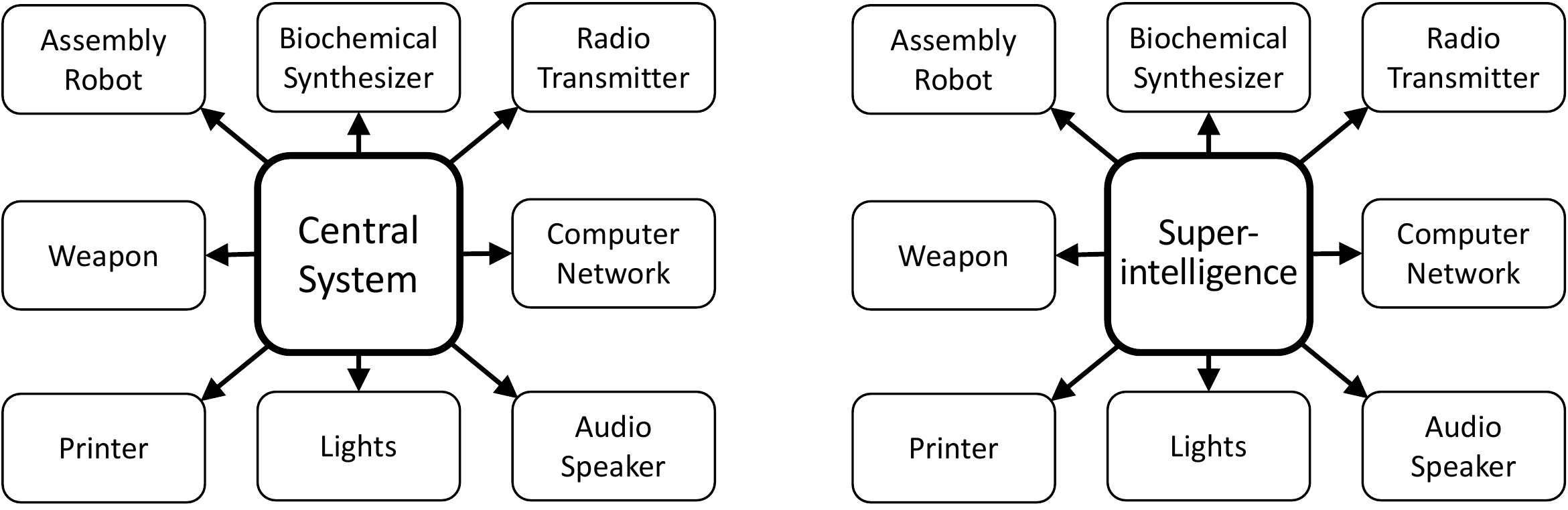}
   \caption{\scriptsize Central systems and peripheral systems. {\sl Left:} A central system has some degree of intelligence and nominally drives one or more peripheral systems to which it is connected (surrounding boxes). A peripheral system physically or communicatively alters the physical world. The peripheral systems shown are examples; a central system need not be connected to all of them. {\sl Right:} Great concerns have been expressed about a superintelligence as the central system.  Note that a superintelligence initially has access to only pre-existing peripheral systems.}
   \label{fig:systems_with_examples}
\end{figure}

Thus, if one is concerned about risk from intelligences, it is necessary to distinguish a ``central system,'' having some degree of intelligence, from the ``peripheral systems'' that the central system uses as effectors to alter the world, as in Figure~\ref{fig:systems_with_examples}.  Because words can alter the physical world\footnote{Example:\ Thomas Paine's pamphlet, {\sl Common Sense}, helped instigate the American Revolution in 1776.} and because markedly impaired word-production is part of the locked-in syndrome, words and communication are considered a peripheral system.

Readers may usefully conceptualize central systems as ``agents'' and peripheral systems as ``tools.''  However, the systems terms are preferred for their emphasis on information flow, and to distinguish from other meanings of ``tool'' [Dyson].

\section{Misplaced Focus on Central Systems}

In contemplating the dangers arising from machine intelligences, scholars to date have focused on characteristics of the intelligence itself, i.e., the central system.  The clearest example is Professor Nick Bostrom's 415-page book whose simple main title -- {\sl Superintelligence} -- clearly signals its focus [Bostrom].

It is straightforward to demonstrate that existential catastrophe for humankind rests in the peripheral systems, not the central systems.  An asteroid of the dinosaur-killing class is a peripheral system with a zero-intelligence central system.  So, too, is the smallpox virus, which undid the Aztec civilization [Oldstone], and the Black Death, which caused demographic collapse across Europe [Kelly].\footnote{George R.\ Stewart's unforgettable 1949 novel, {\sl Earth Abides}, graphically shows how a pandemic of less than universal lethality may collapse civilization [Stewart]. It also exemplifies the ``boring apocalypse'' typology, in which an interplay of multiple critical systems enhance the deleterious effect of the primary insult [Liu et al].}

It is also clear that central systems of modest capacity are extremely dangerous when teamed with powerful peripheral systems. The brain of Joseph Stalin was not superintelligent, but because it was coupled to a powerful far-reaching peripheral system -- the state security apparatus of the Soviet Union -- it could kill one million people in the Soviet Union during just two non-war years, 1937-1938 [Ellman]. Stalin's brain could not have done so if it were in Switzerland.  As further examples, the humans and computers that constitute the central system controlling nuclear weapons do not rise to superintelligence levels.  And, of course, every war in human history has been prodded along by words (a peripheral system) emanating from less-than-superintelligent human brains.

\begin{figure}[htbp]
   \centering
   \figsizeW{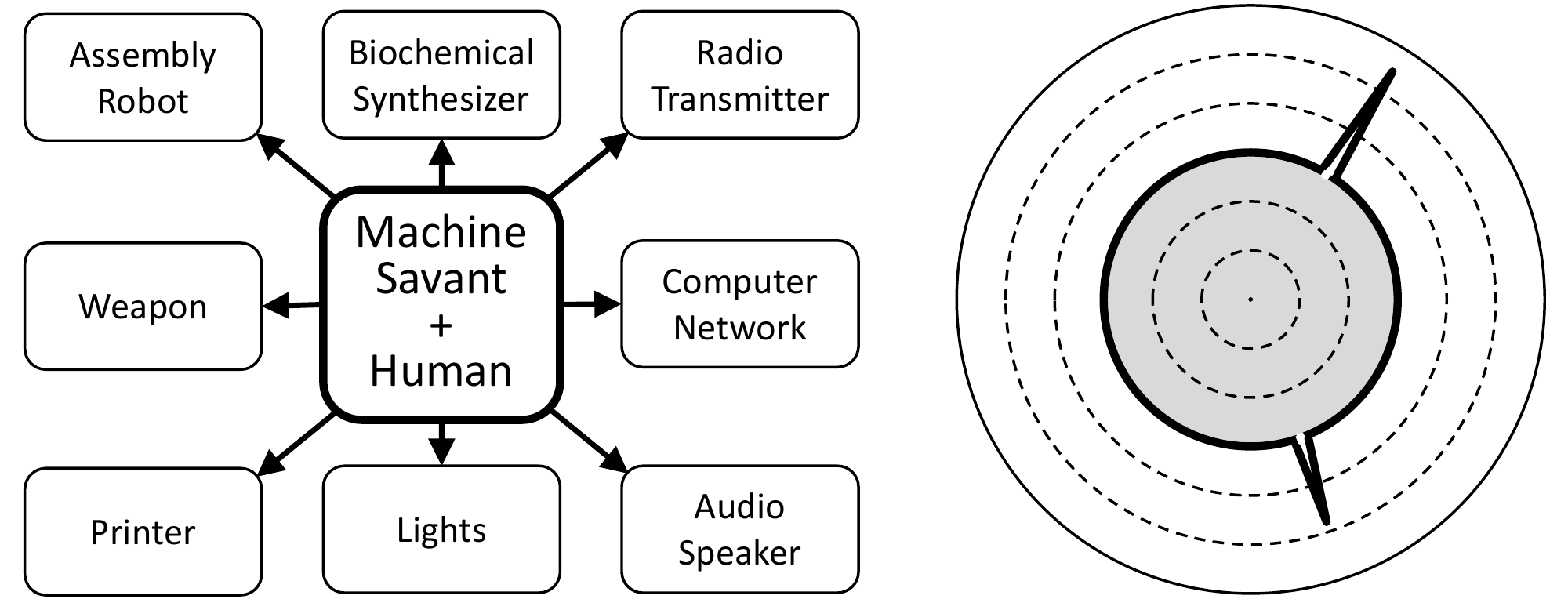}{.8\hsize}
   \caption{\scriptsize A computer-enabled world without machine superintelligence. {\sl Left:} Central systems and peripheral systems. Although the peripheral systems shown are again examples, humans who lead governments or large military organizations may indeed connect to numerous peripheral systems. {\sl Right:} Radar diagram of a notional central system composed of a human and a machine savant, using the same notation as Figure~\ref{fig:radar}. In all domains this combined system functions at least as well as a human, except in the two unnamed domains where it operates supra-humanly.}
   \label{fig:systems_with_human}
\end{figure}

Thus, when scholars interested in machine intelligence ask the question ``What harms can a superintelligence cause?''\ they are neglecting a far more pressing question: ``What harms can result from a central system that is composed of a human intelligence combined with a less-than-superintelligent machine?'' (Figure~\ref{fig:systems_with_human}). Or, more pointedly: What harms can result from intelligent human malevolence teamed with machine savants?  The answer, as we have seen, is simply: ``It depends on the peripheral system.''

\begin{figure}[htbp]
   \centering
   \figsizeW{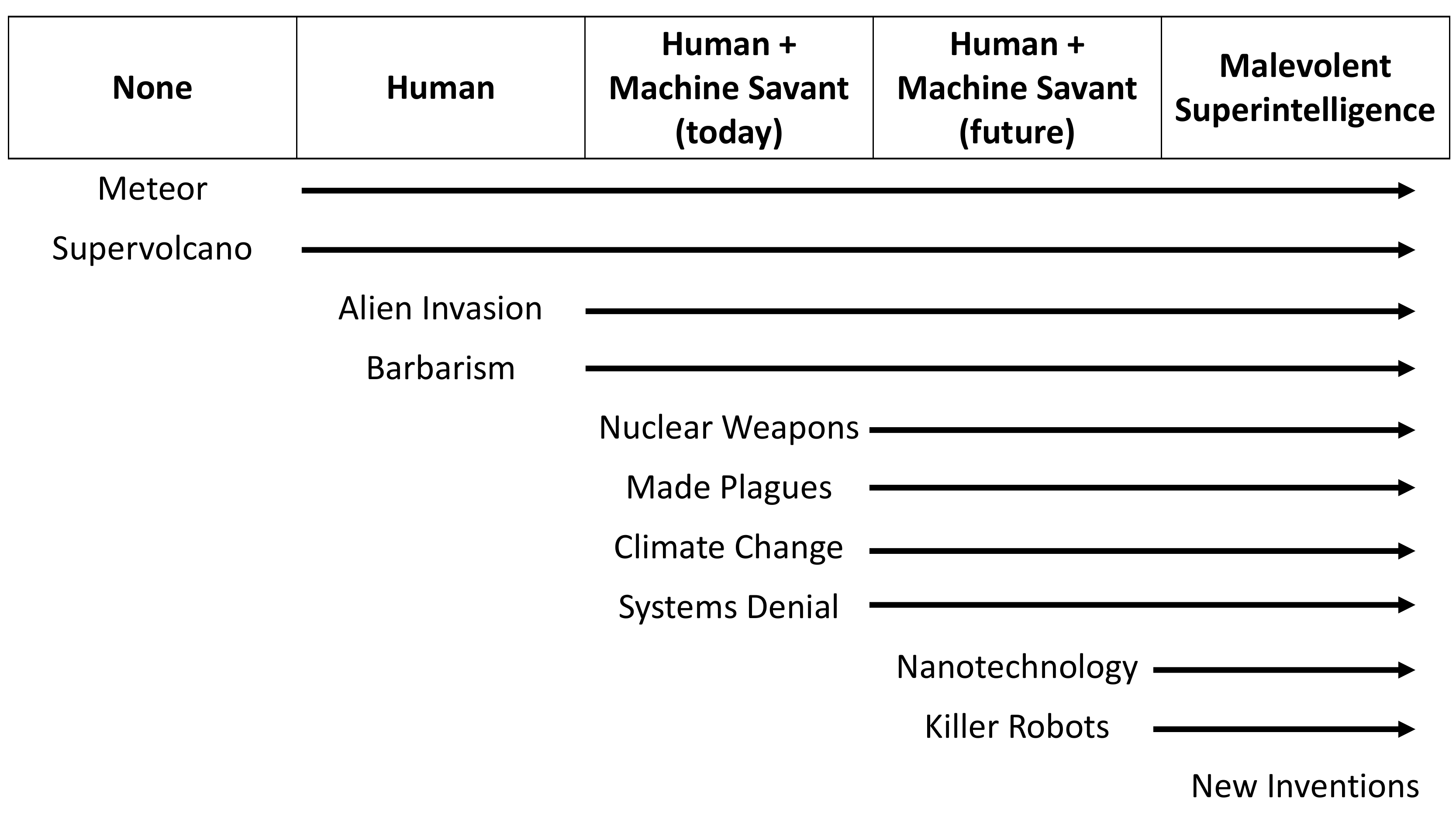}{.605\hsize}
   Time $\Rightarrow$
   \caption{\scriptsize Appearance on earth of peripheral systems that pose an existential threat to humankind (bottom part), as a function of extant central systems (top part, in boxes). Arrows indicate persistence of each peripheral system. ``Alien invasion'' refers to the possibility of attack by a superior extraterrestrial intelligence that became aware of humankind's existence from radio broadcasts before the computer era. ``Barbarism'' encompasses the strictly social causes of the Dark Ages that have collapsed human civilization in the past. ``Systems Denial'' is the putative inability of today's civilization to function without internet-connected systems, typified by the paralyzing ransomware attacks on Maersk in 2017 [Greenburg] and Colonial Pipeline in 2021 [Nakashima et al]; this risk is debatable, since human civilization existed satisfactorily in 1995 without global internet dependence. It is assumed that ``gray goo'' nanotechnology and killer robots, if they appear, can be made without the participation of a superintelligence. The duration of the ``Human + Machine Savant (future)'' period is unknown.}
   \label{fig:threats2b}
\end{figure}

This realization leads to a second conclusion that further diminishes the near-term relevance of machine superintelligence: For a given peripheral system, every threat that includes a superintelligent central system is duplicated or presaged by a similar threat in which the central system combines a machine savant with one or more humans
(Figure~\ref{fig:threats2b}).

As a concrete example, Bostrom [pp119-120] describes a malevolent superintelligence intending to destroy humankind by plague.\footnote{``Malevolent'' and ``intending'' reflect a human victim's assessment of the superintelligence and its actions. Although the actions may be driven by mechanisms as banal as a misaligned utility function -- arising intentionally or unintentionally -- combined with backward-chaining, they certainly have the {\sl appearance} of malevolence and intention. Thus, this work uses ``malevolent'' synonymously with ``harmful.''
} In his scenario, the superintelligence designs a microbe to accomplish this goal, then coopts and/or hoodwinks human accomplices to procure the necessary supplies and to synthesize the organism.  Although this is a perfectly valid and internally consistent scenario, it would be misguided to worry about it today. The more proximate concern is that malevolent humans -- which already exist in abundance -- will use savant-level software to design an ``end-times'' microbe, and will use their own resources (personnel and lab control software) to synthesize it with existing techniques. Part~II will demonstrate that the microbial threat is indeed proximate.

\begin{figure}[htbp]
   \centering
   \figsizeH{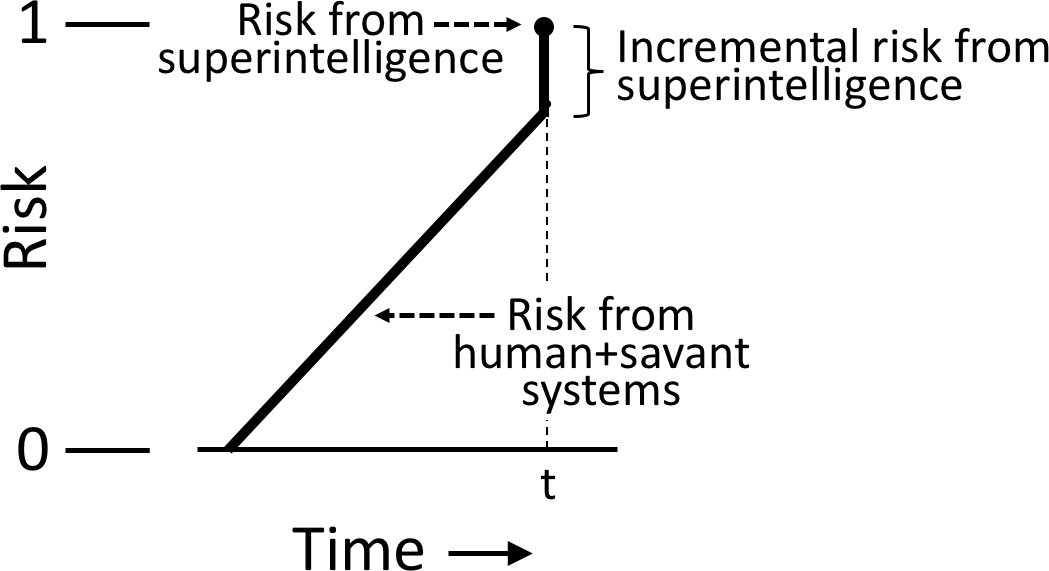}{1.5in}
   \vskip -16pt
   \caption{\scriptsize Hypothetical example of small incremental risk from a superintelligence, despite high absolute risk. The existential risk from combined human-savant central systems rises continuously over time, reaching a very high level. At time $t$ a superintelligence appears, almost instantaneously bringing a nearly 100\%\ existential risk to humankind. However, despite this high absolute risk, the risk to humankind rises little at time $t$ because of a ceiling effect: risk cannot exceed 1, and there is little space to advance from the already-high human+savant risk. Today, the shapes and dimensions of both the human+savant curve and the superintelligence curve are unknown, but the present work argues that enough is known of the curve for biotechnology, as controlled by humans, for it to warrant preferred attention well ahead of speculations on superintelligence.}
   \label{fig:incremental}
\end{figure}

Thus, in gauging the existential risk from the biotechnology peripheral system, a superintelligence in the central system offers little additional risk because of the risk derived from human-plus-savant central-systems that occurs earlier in time (Figure~\ref{fig:incremental}). Any sensible mitigation strategy should, therefore, focus on the human-derived risk ahead of the speculative superintelligence-derived risk, assuming the human-derived risk is significant.  Or, more colloquially, just as ``Airway'' traditionally comes before ``Breathing'' and ``Circulation'' in cardiopulmonary resuscitation, humans should get their own house in order before worrying about superintelligences.

Putting the focus back on peripheral systems, the instant a machine superintelligence takes shape, it will have at its disposal several peripheral systems by which to extinguish human civilization.
The rise of a superintelligence, therefore, merely adds another entity to the list of sentient intelligences who are capable of pulling the end-times trigger, as \autoref{section:math} and [Torres] explore in detail.  The superintelligence may choose to invent new peripheral systems (last line of Figure~\ref{fig:threats2b}), but that would be unnecessary for its human-killing goals if humankind had left pre-existing end-times peripheral systems unmitigated, ready for the just-awakened superintelligence to use effectively (the earlier lines of Figure~\ref{fig:threats2b}).

As an example, over many decades the United States has labored to create precisely such mitigations for control of its nuclear weapons [Schlosser], with the aim of creating a ``fail-safe'' peripheral system that prevents any unauthorized human from initiating a nuclear strike. Inevitably flawed, but continually refined, these mitigations have so far worked. By contrast, many civilization-sustaining peripheral systems -- producing and moving food, water, energy, and information -- are connected to the global internet without similar mitigations, enabling central systems consisting of national leaders, human criminals, and even ``script kiddies'' [Putman] to unleash breakdowns of unplumbed depth.  If these peripheral systems are so exploitable that 48 billion robocalls in America each year cannot be prevented [Palmer], then our destruction need not await a machine superintelligence.

A robust approach to mitigating existential risk assumes the worst of the central system and concentrates on limiting the destructive outcome of peripheral systems. That way, a Stalin-oriented superintelligence, newly awakened, finds itself among peripheral systems having a misuse potential more like Switzerland than the Soviet Union.

\section{Mistaken Hopelessness}

According to [Bostrom p116], the overwhelmingly superior intellectual capability of a superintelligence -- perhaps as superior to humans as we are to beetles [Bostrom p112] -- will enable it to foresee and evade all defenses humans contrive against it. One may believe, therefore, that all work to strengthen peripheral systems against misuse by a superintelligent central system would be wasted effort, because it is doomed to fail.

This is a dangerous belief because, first, it could lead to hopelessness and inaction at a critical point in human history, and second, because it is wrong.

It is wrong for the obvious reason that a superintelligence may never appear -- for whatever reason -- whereas malevolent central systems composed of humans plus savant software exist now. Moreover, extraordinarily capable savant software for biotechnology {\sl must} come into existence before superintelligent software does, because such savant software is a mandatory component of a superintelligence that is definitionally required to excel in ``all domains of interest.''\footnote{Attempting to neutralize savant-associated risk, by urgently creating a beneficent superintelligence to protect the human race, would be a literal {\sl deus ex machina}, i.e., the creation of an infallible, immortal, all-knowing intelligence, ultimately with the ability to create souls and tend them [Bostrom pp122-123]. Although building God would solve many problems, as the pinnacle of hubris it cannot be recommended as a realistic solution to anything.} Failing to limit the destructive potential of peripheral systems exposes humankind to very real existential threats.

Hopelessness is also wrong because mitigating the threat from peripheral systems may indeed succeed against a superintelligence, given that (a) a superintelligence is not required to be all-seeing, and (b) invulnerability need not develop simultaneously with superintelligence.  For the latter case, consider a machine that becomes superintelligent in just minutes, and immediately devises an impregnable plan to foresee and resist attacks by humans. However, what if {\sl executing} this plan requires some item not immediately available to the machine, e.g., seven billion doses of vaccine-resistant smallpox, or the DNA sequence of every human, or 5G transmitters spaced every 1000 feet worldwide? No matter the brilliance of the superintelligence, these physical items simply cannot come into being for some time. This creates a vulnerability window in which it remains possible for human preparation to frustrate the superintelligence's plans. Limiting the tools available to the superintelligence is, therefore, a wise strategy.\footnote{The logical extreme of this reasoning is clearly true: the best defense against superintelligence is a material civilization incapable of building it.}

This should not surprise: intellect is not all.  Every year, tiny-brained snakes succeed in biting 5 million big-brained humans, and one hundred thousand humans die [Warrell].

\section{Quantifying the Risk of Superintelligence}\label{section:math}

If a newly-arisen superintelligence is viewed as an increment to the number of sentient entities that can extinguish civilization via some peripheral system, then it becomes conceptually straightforward to quantify the existential risk attributable to superintelligence.

Consider some number, $N$, of independently-acting sentient entities who can end civilization. These entities correspond to central systems.  Assume, further, that all of these entities have the same probability, $P$, of ending civilization in any given year.  Starting from Equation~\ref{eqnBasic} in Appendix~A, [Sotos] showed that the median lifetime in years of this civilization, i.e., the 50th-percentile ``lethal duration'' ($LD_{50}$), is:

\begin{eqnarray}
LD_{50}=\frac{ln(1-0.50)}{N\ ln(1-P)}\label{eqnLD50}
\end{eqnarray}

\noindent So, for example, if $N=3000$ molecular biologists, with each having $P=$ one-in-a-million chance of collapsing civilization each year, then the predicted median lifespan for earth's civilization is 231 years.  If 30,000 molecular biologists exist, then the median lifespan is 23 years.  (Herein, only molecular biologists who know how to collapse civilization ``count.'' Thus, the term is subtly different from the everyday meaning of ``molecular biologist.'' Even among this group, $P=10^{-6}$ per year is probably too low.)

When a superintelligence appears, $N$ increases by one. We initially assume it has the same $P$ as the molecular biologists.  Recalculating the median civilizational lifespan using $N+1$, and comparing that to the lifespan with $N$ entities, yields a difference (loss) in civilizational lifespan, which is the existential risk attributable to the superintelligence:

\begin{eqnarray}
\Delta LD_{50}&=&\frac{ln\ 0.5}{N\ ln(1-P)} - \frac{ln\ 0.5}{(N+1)\ ln(1-P)}\\[15pt]
&=&\frac{ln\ 0.5}{N\ ln(1-P)}\frac{N+1}{N+1} - \frac{ln\ 0.5}{(N+1)\ ln(1-P)}\frac{N}{N}\\[15pt]
&=& \frac{ln\ 0.5}{N\ (N+1)\ ln\ (1-P)}\label{orderNN}
\end{eqnarray}

\noindent In the 3000-molecular-biologist example, adding the 3001st entity, whether machine superintelligence or human, shortens the civilization's median lifespan by 28 days.  In the 30,000 example, the 30,001st entity shortens the civilization's lifespan by almost nothing: less than 7 hours.

The $N^2$ term in the denominator of Equation~\ref{fig:threats2b} means there will be little risk effect -- perhaps even trivial -- from introducing a superintelligence into a world where a large number of humans can end civilization.  Thus, the world's first superintelligence, technologically momentous as it will be, could existentially be no more significant than adding a 3001st or 30,001st molecular biologist, if $P$ is uniform.

The assumption of uniform $P$ is more general than it may seem.  Appendix~A shows that, so long as no entity has $P$ larger than about 0.1, simple arithmetic can recast any scenario having non-uniform $P$ values into an equivalent scenario having uniform $P$ values and the same $LD_{50}$, but a different $N$.  For example, an entity with thrice-normal $P$ would be recast to three entities having normal $P$. Thus, even the creation of a superintelligence having a $P$ that is 100 times larger than the uniform $P$ of human molecular biologists would be risk-equivalent to the creation of molecular biologists 3001 through 3100\ ...\ hardly a newsworthy event. Going forward, references to ``uniform $P$'' will include well-behaved scenarios that have been transformed in this way.

We may now consider a machine superintelligence that is massively more murderous than humans.  Scholars in the existential risk community in 2008 offered 5\%\ odds that a machine superintelligence (S) would kill a billion humans before the year 2100, i.e., 0.05 over 92 years [Sandberg and Bostrom]. To be generous, we can set $P_{S}$ much higher -- 0.05 {\sl per year} -- so that $P_{S}$ is 50,000 times greater than our baseline $P=10^{-6}$ for molecular biologists.  Hence, by Equation~\ref{eqnLD50approxvariable} in Appendix~A, the appearance of this highly malevolent machine superintelligence is equivalent to bestowing on 50,000 humans the capability to end civilization, each with one-in-a-million annual odds. However, as will be shown later, a scenario with 50,000 such humans is both more immediate and more probable than the rise of software that wants to kill us all.

Extensions to Equation~\ref{orderNN} would show a ceiling effect, per Figure~\ref{fig:incremental}.  For example, if the existence of 3000 morally unconstrained molecular biologists yields a 99\% chance of ending civilization in the next 20 years, then the addition of a murderous machine superintelligence can maximally add only a 1\% risk in that time frame.  No superintelligence can bring more risk than humans leave open.

Equation~\ref{eqnLD50} encapsulates one more arithmetic insight.  Assuming uniform $P$, merging {\sl all} the peripheral systems of Figure~\ref{fig:threats2b} into a single class, so that $N=N_{\tt nuclear\_weapons}+N_{\tt made\_plagues}+...$\thinspace, demonstrates that the overall risk to humankind's future is dominated by the peripheral system having the most entities that can trigger it (because $N=N_{\tt most}+N_{\tt leftovers} \approx N_{\tt most}$).

Recent and continuing progress in biotechnology arguably positions made-plagues as the current ``most'' system. Thus, if new savant-level software, far below the level of superintelligence, substantially increases $N_{\tt made\_plagues}$ (i.e., $N_{\tt most}$), that would bring immediate, substantial existential risk to humankind and likely qualify as the most dangerous software development on earth.  Part~II, below, expands this point.

\part{Biotechnology}

Part I observed that the risk from a machine superintelligence can be pre-empted by the risk from humans and savants, and can be eclipsed when a large number of humans direct a civilization-ending peripheral system.

Part II examines this ``large number'' through the lens of biotechnology, which has emerged as a peripheral system with substantial potential to end civilization via made plagues. Unlike nuclear weapons, biotechnology has enormous societal support and does not require a heavy industrial infrastructure; it may, therefore, soon be wielded by a large number of individual humans. Part~II also illustrates how the other component in a human-containing central system, savant-level software, can stimulate proliferation by making biotechnology easier to use, hence accessible to more people.

Of note, in assessing the risk attributable to central systems composed of humans and machine savants, the worst possible actions from the human must always be anticipated. This is true even if the human intelligence must operate at the highest levels, because high intelligences may become malevolent or diseased,\footnote{The case of John Nash is especially sobering. His mathematical descriptions of human economic behavior earned him a Nobel Prize, but his ``beautiful mind'' later harbored severe schizophrenia [Nassar].} and because humans may excel in one domain, yet fail utterly in a moral domain.  Thus, human involvement in the central system can never be relied on to brake the path to disaster.

\section{Biotechnology as a Dual-Use Peripheral System}

The laudable goal of reducing human sickness animates the vast majority of biotechnological development. The field started in 1972, with the first controlled construction of a novel DNA molecule [Berg and Mertz] [Cohen].  Since then, biotechnology has revolutionized the treatment of heart attacks, stroke, cancer, auto-immune disease, and many other conditions. It enabled the rapid development of covid-19 vaccines in 2020 and is poised to eliminate one of humankind's great scourges -- malaria -- by using a ``gene drive'' to cause extinction of the Anopheles mosquito that transmits the disease [Hammond \&\ Galizi].

However, equally rational viewpoints would classify humankind's biotechnology in 2021 as an existential threat to multiple Anopheles species and as a genocidal weapon against cancer cells.  These alternate viewpoints perfectly illustrate the danger that is unalterably inherent in biotechnology: it is dual-use [Hoffman p 132], meaning it yields weapons as well as medicines.  One organism's biotech salvation is another's biotech nightmare.

Today, great controversy surrounds the question of whether humans deliberately created the covid-19 virus, by adapting it from a natural coronavirus to become deadlier and more contagious [Anderson et al] [Wade]. The most important facet of the controversy is not the answer to the question, whatever it may be, but the absence of claims that humans in 2019 were incapable of engineering such a virus.

That lack of skepticism, alone, is a Rubicon for humankind's future: it means that today there is no doubt that a cadre of humans already has the skills to create world-girding microbes of unnatural virulence.

Whether biotechnology becomes a nightmare or a salvation for Homo sapiens depends directly on choices that humans will make. Despite thousands of good choices, just one bad one could bring the nightmare.

\section{Numberings}

Figure~\ref{fig:threats2b} shows the time horizons for several existential threats. For two of the eight threats that exist today -- nuclear warfare and engineered plagues -- the actions of one properly resourced individual human can trigger a global catastrophe. While the putative number of such humans is stable and small for the nuclear threat (n=2), the situation is entirely different for the biotechnology threat.

In biotechnology, we might assume that anyone having sophisticated knowledge of genetic techniques is able to create, or at least duplicate, an end-times plague.  Of course, there is no clear definition of ``sophisticated knowledge,'' but as one proxy, from 2008 through 2015 approximately 180,000 individuals authored 5 or more articles in the scientific literature having the term ``genetic techniques'' as a major keyword [Sotos].  From Equation~\ref{eqnLD50}, if 180,000 persons each had an annual one-in-a-million probability of creating an end-times plague, our civilization's median lifespan would be 3.8 years. That is nightmare enough, with no need to postulate script kiddies branching into biotechnology.

It may be argued, however, that scientific knowledge alone is insufficient to collapse civilization, and that significant capital resources will be required -- to hire personnel and buy the equipment needed to develop and deploy a civilization-ending microbe.  In 2020, the entire annual budget of the U.S.\ National Institutes of Health was \$42 billion.  Even if it took \$50 billion to hire the scientists and buy the equipment, two dozen people and families on earth have such resources today; if it took \$10 billion, then 228 people and families have the resources; if \$1 billion, then 2755 people and families have the resources...\ including the Kardashians [Dolan et al].\footnote{As far back as 2014 the author knew a PhD molecular geneticist who thought that \$1 billion would enable someone with his level of training to kill 95\%\ of humankind.}  From Equation~\ref{eqnLD50}, if 2755 entities each had an annual one-in-a-million probability of creating an end-times plague, our civilization's median lifespan would be 251 years. If one of them had a 15\%\ chance, then it would be 4.2 years.

Bad decisions could enormously increase proliferation.  For example, the CEO of Twist Biosciences aims to use DNA molecules as an information-storage alternative to disk drives [Wired]. This would seem to require the ability to synthesize long strings of {\sl arbitrary} DNA at very low cost. Her cost goal is less than 8~cents for a DNA string having the size of the entire human genome. Extrapolating, it would therefore cost 0.004 cents to manufacture the DNA for a smallpox virus (the sequence is available on the internet). Presumably, a trivial looping script could print essentially unlimited copies. If Twist applies for patents, their underlying technology would become public, allowing others to remove any safeguards that Twist might have built in.

This section has been necessarily speculative.  No one knows (or is saying) exactly what needs to happen before the genie is freed completely from the bottle.
Humans first created a synthetic life form in 2010 [Gibson et al].  In 2012 two research teams deliberately enhanced the contagiousness of an influenza virus [Herfst et al] [Imai et al].  In 2001 a team enhanced the lethality of a close relative of smallpox [Jackson et al]. In 2016 a team synthesized a different close relative of the smallpox virus from pieces of DNA they obtained by mail order [Kupferschmidt], leading them to conclude that ``no viral pathogen is likely beyond the reach of synthetic biology'' [Noyce]. Yet, we still inhabit biotechnology's childhood, which was born less than one working lifetime ago.

\section{Biotechnological Machine Savants}

If a human chooses to design, build, and deploy a biotechnological weapon, how might machine savant software make that effort easier, thereby making it accessible to more people?

The answer is best explained if the weaponeer's fundamental challenge is recast as a {\sl search} problem. As an example, consider the SARS-CoV-2 virus, which, although not an end-times plague, has devastated the world.  Conceptually, it is merely a very large number.  Figure~\ref{fig:genome} represents the SARS-CoV-2 genome as a single hexadecimal number
-- it fits on one printed page.  This {\sl number}, when translated into RNA and inserted inside a human cell, becomes integrated into the cell's metabolic processes and yields polymer and other molecules that assemble themselves into what we recognize as a virus (Figure~\ref{fig:virus}).

\begin{figure}[htbp]

\def\Z#1{#1\hskip 0pt}
\advance\baselineskip by -17pt
{\fontsize{3}{3.5}\selectfont\noindent\input{fignum60_genome.tex}}
\caption{\scriptsize Genome of the SARS-CoV-2 virus as a 14,952-digit hexadecimal number, with each digit representing two base pairs. Each RNA base was encoded as A=0, C=1, G=2, U=3. This number killed 4 million humans in 18 months. Downloaded April 27, 2021 from: \url{https://www.ncbi.nlm.nih.gov/nuccore/NC\_045512} \ \  Compared to previously known coronaviruses, changes in 17 of these digits/base pairs were the keys to the emergence of SARS-CoV-2 as a pandemic virus [Anderson et al].}

\label{fig:genome}
\end{figure}

\begin{figure}[htbp]
   \centering
   \figsizeH{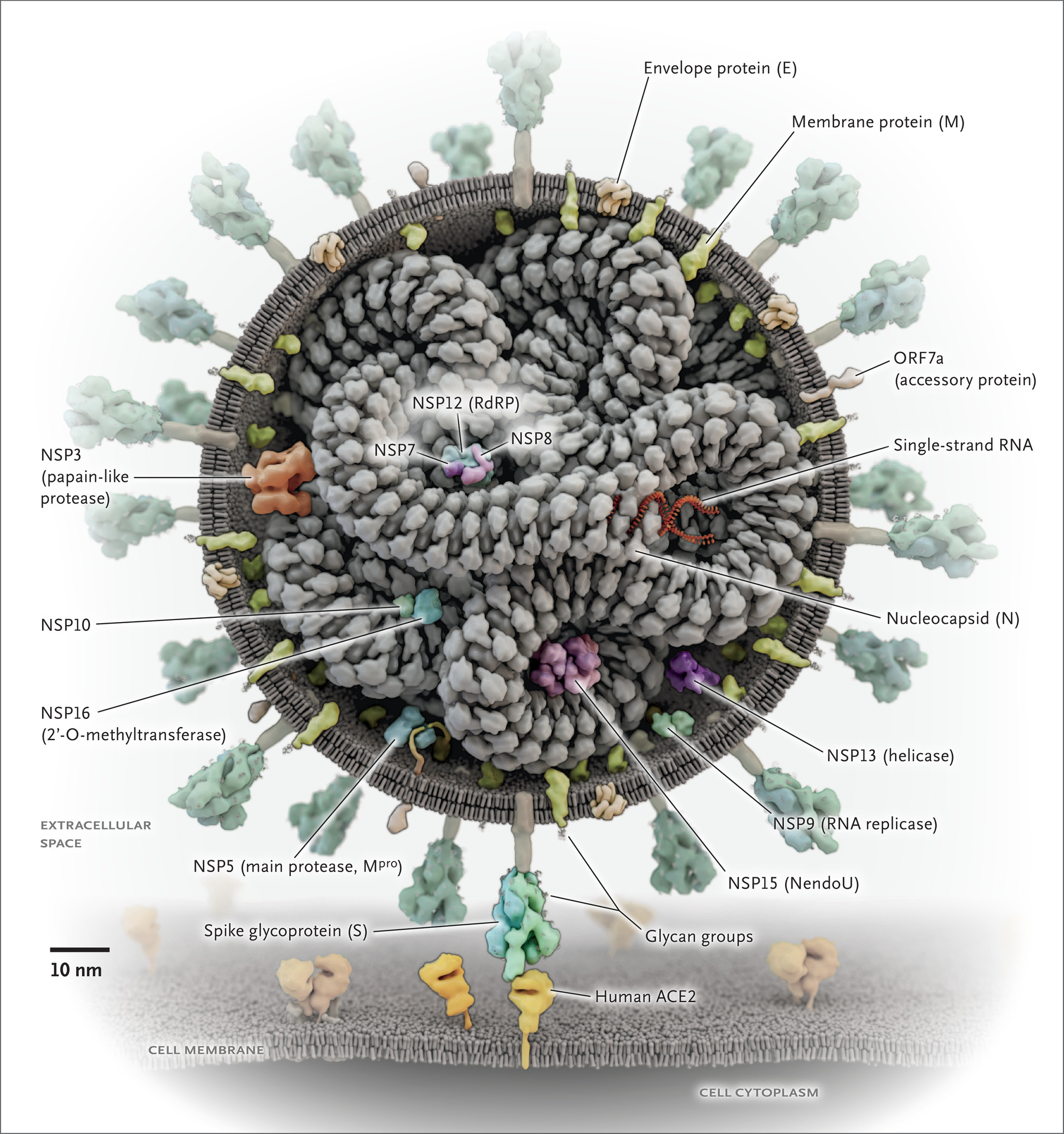}{.9\vsize}
   \caption{\scriptsize When the number from Figure~\ref{fig:genome} is integrated into the metabolic processes of a human cell, this dangerous object -- the SARS-CoV-2 virus -- is created. It can self-replicate with great efficiency. Reproduced with permission from  [Parks \&\ Smith], Copyright Massachusetts Medical Society.}
   \label{fig:virus}
\end{figure}

Importantly, the nearly 30,000 base pairs of the SARS-CoV-2 genome differ in only 17 significant ways from the genome of a naturally occurring bat coronavirus [Anderson et al].  If aspiring but unsophisticated weaponeers started with the bat virus, and knew only that 17 changes were required to transform it to a weapon, they could adopt a brute force method to identify the weapon by {\sl searching} through all variants of the bat virus having 17 differences. Because this ``search space'' would include an astronomical number of variants, the brute force search method is not practical.

In general, therefore, the challenge of designing a bioweapon reduces to the broader task of finding methods to narrow a search space -- a general task that has long been at the leading edge of computer science [Barr \&\ Feigenbaum].

With that recasting, we next examine a few intersections of software and biotechnology to see how software might accelerate the creation of bioweapons.  The examples are chosen only for clarity, and are of course incomplete.

\subsection{Replicating Human Expertise in Software}

Today, if a human were to weaponize a virus -- for example, the influenza virus -- that human must possess a deep understanding of influenza virus biology. Otherwise, the virus' genome is so large that making random, uninformed changes in it would be statistically unlikely to yield a virus with the highly refined (but awful) characteristics of an effective weapon.

Such expertise is scarce.  Endowing a machine savant with the biological knowledge of a human influenza virus expert would be an obvious step in proliferating, and thereby accelerating, the discovery of both influenza treatments and weapons.   Moreover, from an engineering standpoint, this would be far less challenging than developing a general artificial intelligence or a machine super-intelligence, given that software architectures to transfer expert knowledge have been available for decades [van Melle et al].

\subsection{{\sl In Silico} Modeling}

{\sl In silico} modeling of biological and biochemical systems has emerged as a powerful and widespread tool to discover new treatments for diseases.  For example, early in the covid-19 pandemic, Oak Ridge National Laboratory unleashed the world's most powerful computer for 24 hours to run models predicting the binding of one billion different chemical compounds to proteins in the SARS-CoV-2 virus [Parks \&\ Smith] [Acharya].

In other words, the computer rapidly traversed an initial search space of one billion treatment candidates, eliminating those predicted to have low anti-virus efficacy by virtue of their poor binding to the virus.  This left a much smaller search space of candidates to be tested in the slow, subsequent steps of clinical testing. No human could have performed this savant-like task with brain alone -- the models are far too complicated [Naderi].

Similarly, and unsurprisingly, pharmaceutical companies have invested in computer systems that predict when a particular candidate drug molecule will cause unwanted, market-killing side effects [Wang].  Perversely, but in keeping with the dual-use nature of biotechnology, these systems also contain the knowledge needed to deliberately create molecules that {\sl cause} side effects, i.e., weapons.

In general, computer models provide a means to perform experiments at a scale that humans cannot, thereby narrowing the search space.

\subsection{Software-Controlled Laboratories}

Biotechnological danger from completely unremarkable computer software is also considerable.  Computer control of industrial-scale wet-ware laboratories (e.g., [Chory], [Strickland]) could leverage evolutionary selection to produce microbes having specific characteristics.  Here, the difficult tasks are in the laboratory functions, not the computer control, but it is the automation that delivers the scale to rapidly reduce the search space by screening enormous numbers of candidate microbes.

One could therefore imagine a large computer-controlled laboratory aiming to evolve a variant of the Ebola virus that spreads through the air.\footnote{This idea is from [Clancy p425], although airborne transmission is entertained with any new pathogenic virus.  In the 1980s the Soviet Union tried to build robotic laboratories that would create a new viral weapon each month [Hoffman p107]. Today, they might succeed.} Ominously, doing so is only a matter of will and resources, because this general evolutionary approach is well established: a 2018 Nobel Prize was awarded for developing ``directed evolution of enzymes'' with bacteria as factories [Arnold].\footnote{Even earlier, in the 1940s, evolutionary techniques were used to increase penicillin yield from molds [Hobby].} Lab-on-a-chip technology, supplemented by the emerging organ-on-a-chip technology that could include human organoids [Azizipour et al] [Schutgens \&\ Clevers] [Kim et al], will enable complex, highly parallelized, biological evolutionary selection functions, thereby bringing directed evolution to human-microbe interaction.

Here, a comparison with machine super-intelligence is instructive.  One of the pre-eminent concerns about machine super-intelligence is the possibility that a moderately advanced digital intelligence could design a more intelligent digital successor that would quickly design a still more intelligent digital successor and so on, leading to an explosion of digital intelligence that could be malign [Bostrom pp75-94].  As yet, these concerns remain hypothetical, because the engineering basis for such an explosion remains unknown.

By contrast, directed-evolution-in-a-laboratory already provides an exact, actionable blueprint to create and discover biological malign entities while leveraging the short time scale of microbial generations (for SARS-CoV-2, on the order of 10 hours [Bar-On et al]). It would simply be a speedier reimplementation of nature's proven technique that has produced all the scourges, known and unknown, that afflict all the living things on earth.

\subsection{Software-Assisted Tools}

Although scientific progress historically arises from the arrival of both new paradigms and new tools, in recent decades tools have become pre-eminent [Dyson], especially in biology [Jogalekar]. Without doubt, all parts of biology have benefitted -- and will continue to benefit -- from tools that incorporate computer software.  It is therefore reasonable to believe that advances in software will spur further advances in molecular biology -- and thus the potential for bioweaponry.

Synthetic biology, long considered an existential threat [Block], but also ``a powerful tool to create therapeutics which can be rationally designed'' [Leventhal et al], is a case in point.  As a specific example, in discussing the prospects for designing synthetic ``oncolytic'' viruses (OVs) to treat cancerous glioma cells, Monie and colleagues highlight no fewer than 10 software-enabled capabilities within systems biology that enable rational synthetic biology:

\

\begin{addmargin}[1.5\parindent]{1.5\parindent}
\small
Systems biology complements synthetic biology by enabling
complex design and analy\-sis of genetic circuits. Tumor
microenvironment cell states can be defined by using
single-cell multiomics data sets such as RNA sequencing
and mass cytometry. These high-dimensional
data sets are amenable to machine learning and network-based
classification. Computation identifies biological
features unique to glioma cells, elucidates pathways
orthogonal to genetic circuit designs, and helps prioritize
OV-based approaches. After the synthetic OVs are
constructed and tested, further systems analyses ensure
that the OVs have predictable effects on the cells and the
greater tumor microenvironment, statistically accounting
for complex stochastic behavior. This knowledge is fed
back into the design of improved, next-generation OV genetic
circuits. [Monie et al]
\end{addmargin}

\

This is search-space-reduction of the highest order.  If synthetic biology carries existential risk, then Monie et al show that even the comparatively low-performing software inside sequencing and cytometry machines contributes to that risk, as do domain-independent software techniques such as machine learning and statistics.  The larger lesson is that progress and proliferation of software tools, teamed with human intelligence, leads to progress in molecular biology, which in turn enables smarter choices in search space to generate better anti-glioma bioweapons.

Or, in simpler terms, the rising software tide floats all boats, including the warships.

\subsection{Trends}

With biotechnology having passed the Rubicon stage where engineered pandemic organisms are accepted as possible, the central question has become: How easy and how fast will it become to design end-times infectious weapons?

Over the past decades, biotechnology and computer capabilities have risen together.  The brief and limited examples above show that computer software brings a powerful capability into the central system for narrowing search spaces in the design of biological peripheral systems.  Moreover, this capability has been, and will continue to be, achieved with software that functions far below the level of a machine superintelligence.

\begin{figure}[htbp]
   \centering
   \figsizeW{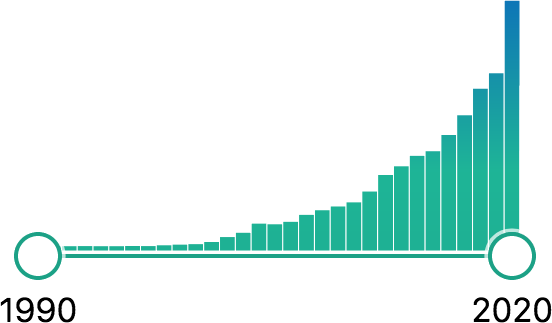}{.35\hsize}
   \caption{\scriptsize Scientific interest in biocomputation. Vertical bars indicate each year's number of publications in the PubMed database that contain the phrase ``in silico'' in the title of the publication. None appeared before 1993 [Sieburg et al]. The total number of such publications is 11,392. Compiled from pubmed.gov on June 30, 2021.}
   \label{fig:insilico}
\end{figure}

It is certain that innovation in biocomputation will continue, and will endow future biotechnological savants with improved power to reduce search spaces. Figure~\ref{fig:insilico} shows how the scientific interest in biocomputation has exploded in less than one working lifetime, as indicated by the annual number of publications in the biomedical literature whose title contains the phrase ``in silico.''

Publication counts do not, of course, predict changes in search-space-reducing capability.  But, just as [Bostrom pp 43, 84] worries that a single breakthrough could deliver a superintelligence far sooner that expected, an unexpected breakthrough in biology (or computer science) could deliver a super-searching machine savant far sooner than the smooth curve of Figure~\ref{fig:insilico} might suggest.  The possibility of such discontinuities cannot be discounted, given molecular biology's past history of wholly unexpected Nobel-honored breakthroughs, headed by the polymerase chain reaction, CRISPR/Cas9, phage display for producing monoclonal antibodies, and RNA interference.

\section{Defensive Savants}

Although vaccines against covid-19 were developed with unprecedented speed, and although physical distancing was a powerful defensive force when used, it should not be assumed that pandemic defense is a solved problem and that all microbes will be susceptible to such interventions.  Had the SARS-CoV-2 virus spread in water, as cholera does; or had it spread in the air over intercontinental distances, as do the microbes that ride the winds from Africa to America [Griffin]; or had there been an environmental reservoir, as with anthrax; or had its clinical syndrome been a usually-mild initial illness followed by years of asymptomatic contagiousness before the fatal culmination, as with HIV; or had it been more difficult to develop a vaccine, as the HIV virus has demonstrated to human science every day for 40 years, the situation would have been incomparably more dire.

Because the threat from biotechnology depends on the gap between ``offensive'' and ``defensive'' biotechnological systems, the need to perpetually develop more capable defensive systems is apparent.  Machine intelligence can certainly assist with defense, but defense has inherent characteristics that limit its ability to keep pace with offense:

\begin{itemize}

\item Time scales in biology greatly favor offense over defense, because microbes can spread widely before countermeasures can be developed and fielded.

\item Defense requires deeper domain understanding than offense because it is harder to repair life than to break it. For example, safety  constrains defense, but not offense.

\item Defense requires broader domain understanding because offense can choose any attack from a large target space, whereas defense must be ready to counter anything.

\item Humans are actively trying to overcome the body's immune defenses, e.g., physicians altering a patient's immune system so it will not destroy a gene therapy virus injected into the patient [Shirley et al] [Thacker et al].

\end{itemize}

\noindent It would seem, therefore, that even with the assistance of biotechnological savants, offense will be first and stronger, and defense later and weaker.

\section{Biotechnology Summary}

Four hundred and eight pages of small print constitute the 1995 book, {\sl Encyclopedia of Plague \&\ Pestilence}, which lists 474 such events in the preceding 500 years, including 165 in the 20th century [Kohn].  Clearly, epidemics are inherent to human existence, antibiotics notwithstanding.

Despite our species surviving all of those epidemics, it is reasonable to hypothesize that some nucleic acid sequence, i.e., a number perhaps not too different from the one shown in Figure~\ref{fig:genome}, codes for an ``end-times'' microbe that would cause a pandemic so severe as to collapse human civilization on earth. Certainly it is not possible to reject the possibility.\footnote{Any rejection attempt that cites some precedent in extant microbes may itself be rejected, as follows. Reproductive fitness has been the sole selection factor for natural microbial evolution on earth. A weaponeer would impose a different selection regime, as Soviet bio-engineers did in the 1980s when they developed a multiple sclerosis weapon that had two clinical phases [Hoffman pp109-110, 298-299]. Even in the natural world, the supposed inverse relationship between virulence and contagiousness has multiple counter-examples, including a single chytrid fungus responsible for the {\sl extinction} of numerous amphibian species [Casadevall], and the delta variant of the SARS-CoV-2 virus, which is both more transmissible and more serious than earlier variants [Sheikh et al].} If one grants the possibility of such a microbe, then civilization survives only because of the enormously large search space of possible genomes for it. This is security by obscurity -- a protection that fails as soon as the key is discovered.

If there is a genomic key-number that can end human civilization, software will aid in its discovery.  Superintelligent software will not do this, because savant software, created and directed by smart but corrupt humans, will do it first.  Machine superintelligence poses a small existential risk because human-controlled biotechnology poses a large risk, leaving little left over.

\part{Conclusion}

The present analysis shows the pitfalls of considering superintelligence risk in isolation.  It is but one threat confronting humankind and can be properly characterized only in the context of other threats and their components.

Part I decomposed existential risk into central systems and peripheral systems, with peripheral systems as the effector of civilizational destruction, and central systems as their creator and director. Because superintelligent central systems do not yet exist, the proximate existential threats to humankind arise from end-times peripheral systems created and directed by hybrid central systems composed of humans and savant-level software.  Part~I showed that the risk attributable to a superintelligent central system decreases rapidly, even to the point of irrelevance, with the proliferation of human-containing central systems able to trigger end-time peripheral systems.  Superintelligences are merely another fallible sentient entity able to trigger end-times peripheral systems against humanity.

Part II showed that the number of humans able to trigger the end-times with biotechnology may be expected to grow rapidly in the near future as biotechnological progress reduces the intellectual and material resources needed to engineer end-times microbes.  Because this engineering effort is inherently a search problem, savant software, operating well below the superintelligence level, will be an essential contributor to biotechnological progress and proliferation. The longer this proliferation continues -- the longer human civilization survives -- the less incremental risk a superintelligence will be able to muster.  Thus, savant software is currently a greater risk to humankind than superintelligent software and, if its progress continues, is likely to remain so.

Scholars posit that an unforeseen breakthrough in artificial intelligence could, at any moment, enable the rapid creation of a superintelligence [Bostrom p84].  However, such a breakthrough confers great risk only if it precedes breakthroughs in biotechnology -- which have been occurring with remarkable frequency.  Paradoxically, computer science breakthroughs on the way to superintelligence could actually {\sl reduce} the threat from superintelligence, if they enhanced biotechnological savant capabilities: this would increase biotechnological risk, and thereby leave less incremental risk available for superintelligence to bring.

The social forces driving progress in biotechnology are immense, irresistible, and lie precisely in areas that will lead to end-times microbes. By contrast, the forces propelling defensive biotechnology are much more modest, despite steps taken in the wake of the covid-19 pandemic.  Defensive biology is also inherently more challenging.  At the extreme, as long as the existential risk from offensive biotechnology remains unmitigated, it could take only one capable, zealous human, acting alone, to render superintelligence completely irrelevant to the future of humankind.

Studies of superintelligence have performed a valuable service by providing a framework and vocabulary to analyze the existential risks associated with advanced software systems.  However, as Schr\"odinger has observed, ``A new discovery is usually overstated and very often formulated with too much detail that wears off later'' [Schr\"odinger p60]. The software posing the greatest incremental threat to humankind is narrow in scope and is associated with the biotechnology peripheral system, not broad in scope and the sole constituent of a central system.  And so, in a world of limited resources that seeks to avoid existential calamity, directing those resources against the most threatening peripheral systems is the rational approach (effectively reducing $N$). An equally urgent, but probably more difficult, priority is to develop social structures that markedly reduce the probability that humans will trigger an end-times peripheral system (reducing $P$).

Enormous human energy has built the $N$-limiting and $P$-limiting strategies that have successfully contained the hair-trigger threat of nuclear weapons for 75 years. That essential work must continue unabated, but it will be dwarfed by the energy required to build and maintain similar constraints on the biotechnological threats already confronting us.
It would be an existential threat, in itself, if countering machine superintelligence today de-prioritized or distracted us from that new essential work.

\ 

\noindent {\bf Acknowledgements.} The author is deeply grateful to Robert Anue, Barry Hayes, Hugh Rienhoff, Tanya Sotos, and Phil Torres for greatly improving this work with their many insightful, thoughtful comments.  All faults remain the sole responsibility of the author. The views herein are the author's, and do not necessarily represent those of the Department of Defense.

\ 

\part{Back Matter}

\section*{Appendix A}

To facilitate mental manipulation, this appendix shows how to simplify mathematical models in which multiple people (``entities'') have different probabilities of ending civilization.

First we model the most basic scenario, in which $N$ independently-acting entities each have a uniform and unchanging annual probability, $P$, of ending civilization.  After $y$ years, the civilization will have a probability, $\mathbb{P}$, of still existing [Sotos]:

\begin{eqnarray}
\mathbb{P} = (1-P)^{Ny}\label{eqnBasic}
\end{eqnarray}

For the scenario in which entities have non-uniform $P$ values, we pick one probability, $P_0$, as reference, and count the number of entities, $N_0$, having that probability. We do the same for all the different probabilities, $P_i$, $P_j$, etc. We can then rewrite Equation~\ref{eqnBasic} to:

\begin{eqnarray}
\mathbb{P} &=& \lbrack\ (1-P_0)^{N_0}\ (1-P_i)^{N_i}\ (1-P_j)^{N_j}\ ...\ \rbrack^y \\[15pt]
ln\ \mathbb{P} &=& y\ ln\ \lbrack\ (1-P_0)^{N_0}\ (1-P_i)^{N_i}\ (1-P_j)^{N_j}\ ...\ \rbrack
\end{eqnarray}

\noindent We are interested in the median lifespan of the civilization, so will set $\mathbb{P}=0.5$ henceforth.  This entails that the number of years, $y$, is the $LD_{50}$ -- the median predicted civilizational lifespan in years -- as defined in [Sotos] and in Equation~\ref{eqnLD50} above. So:

\begin{eqnarray}
ln\ 0.5 &=& LD_{50}\ ln\ \lbrack\ (1-P_0)^{N_0}\ (1-P_i)^{N_i}\ (1-P_j)^{N_j}\ ...\ \rbrack \\[15pt]
LD_{50} &=& \frac{ln\ 0.5}{ln\ \lbrack\ (1-P_0)^{N_0}\ (1-P_i)^{N_i}\ (1-P_j)^{N_j}\ ...\ \rbrack} \\[15pt]
&=& \frac{ln\ 0.5}{ln\ (1-P_0)^{N_0}+ ln\ (1-P_i)^{N_i}+ln\ (1-P_j)^{N_j}+...} \\[15pt]
&=& \frac{ln\ 0.5}{N_0\ ln\ (1-P_0) + N_i\ ln\ (1-P_i)+N_j\ ln\ (1-P_j)+...}
\end{eqnarray}

[Sotos] shows that $ln\ (1-P)\approx -P$ for small $P$ (the error is about 5\%\ for $P=0.1$ and decreases rapidly as $P$ gets smaller). Thus, if we assume that all $P\leq 0.1$, then:

\begin{eqnarray}
LD_{50} &\approx& \frac{ln\ 0.5}{N_0\ (-P_0) + N_i\ (-P_i)+N_j\ (-P_j)+ ...} \\[15pt]
&\approx& \frac{ln\ 0.5}{-(N_0 P_0 + N_i P_i+N_j P_j+ ...)}
\end{eqnarray}

\vskip 3ex \noindent With $P_0$ as a referent, we set $P_i=(k_i\ P_0)$ and $P_j=(k_j\ P_0)$ to yield:

\begin{eqnarray}
LD_{50} &\approx& \frac{ln\ 0.5}{-(N_0 P_0 + N_i k_i P_0 + N_j k_j P_0+ ...)} \\[15pt]
&\approx& \frac{ln\ 0.5}{-(N_0 + N_i k_i + N_j k_j + ...)P_0} \\[15pt]
&\approx& \frac{ln\ 2}{(N_0 + k_i N_i + k_j N_j + ...)P_0}\label{eqnLD50approxvariable}
\end{eqnarray}

\noindent In the situation where all $P$ are the same, all $N_i$, $N_j$, ...\ will be zero, so:

\begin{eqnarray}
LD_{50} &\approx& \frac{ln\ 2}{NP_0}\label{eqnLD50approxuniform}
\end{eqnarray}

Comparing Equations \ref{eqnLD50approxvariable} and \ref{eqnLD50approxuniform} achieves our goal of making plain the effect of non-uniform $P$ values.  It shows that non-uniform $P$ values (Equation~\ref{eqnLD50approxvariable}) equate to a scenario where $P$ is uniform (Equation~\ref{eqnLD50approxuniform}), except that an adjustment in $N$ has been made.  Moreover, the adjustment in $N$ is intuitively easy to understand.  Having one entity with $P_i$ equates to having $k_i$ entities with $P_0$ (recall $k_i = P_i/P_0$).

For example, if, among $N$ entities, a single entity has a $P_i$ value that is twice the $P_0$ value of all the other entities ($k_i=2$), then that entity is simply counted as two entities having $P_0$, so that Equation~\ref{eqnLD50approxuniform} for uniform $P_0$ values is computed using $N+1$ instead of $N$. Or, if three entities have their $P_i$ as six times the $P_0$ value of the other entities ($k_i=6$), then each of those three entities is counted as six entities having the baseline $P_0$, hence $N$ is increased by $3\times(6-1)$ in Equation~\ref{eqnLD50approxuniform}.  Steps like these can be repeated as often as needed to make all $P$ uniform. The sole restriction is that no entity may have a $P$ value greater than about 0.1.

\section*{References}
\parindent=0pt

Acharya A. Supercomputer-Based Ensemble Docking Drug Discovery Pipelinewith Application to Covid-19. {\sl J Chem Inf Model}.2020; 60: 5832-5852. \doiUrl{10.3389/frai.2020.00065}

\ 

Anderson KG, et al. The proximal origin of SARS-CoV-2. {\sl Nature Medicine}. 2020; 26: 450-452. \doiUrl{10.1038/s41591-020-0820-9}

\ 

Arnold, Frances. Innovation by Evolution: Bringing New Chemistry to Life. Nobel Lecture, Dec.\ 8, 2018. \url{https://www.nobelprize.org/uploads/2018/10/arnold-lecture.pdf}

\ 

Azizipour N, et al. Evolution of Biochip Technology: A Review from Lab-on-a-Chip to Organ-on-a-Chip. {\sl Micromachines} (Basel). 2020; 11: 599. \doiUrl{10.3390/mi11060599}

\ 

Bar-On YM, et al. SARS-CoV-2 (COVID-19) by the numbers. {\sl eLife}. 2020; 9: e57309. \doiUrl{10.7554/eLife.57309}

\ 

Barr, Avron; Feigenbaum, Edward A. {\sl The Artificial Intelligence Handbook}. Volume~1. Los Altos, CA: William Kaufman, 1981. Page 28.

\ 

Beard S, Rowe T, Fox, J. An analysis and evaluation of methods currently used to quantify the likelihood of existential hazards. {\sl Futures}. 2020; 115 (102469): 102469-102469. \doiUrl{10.1016/j.futures.2019.102469}
 
 \ 

Berg P, Mertz JE.  Personal Reflections on the Origins and Emergence of Recombinant DNA Technology. {\sl Genetics}. 2010; 184(1): 9-17. \doiUrl{10.1534/genetics.109.112144}

\ 

Block SM. ``Living nightmares: biological threats enabled by molecular biology.'' Pages 39-75 in: Drell DD, Sofaer AD, Wilson GD. {\sl The New Terror: Facing the Threat of Biological and Chemical Weapons}. Stanford, CA: Hoover Institution Press, 1999.

\ 

Bostrom N. {\sl Superintelligence: Paths, Dangers, Strategies}. 8th impression. Oxford: Oxford University Press, 2017. (Initial version published 2014.)

\ 

Casadevall A. The future of biological warfare. {\sl Microbial Biotechnology}. 2012; 5(5): 584-587. \doiUrl{10.1111/j.1751-7915.2012.00340.x}

\ 

Chory EJ et al. Enabling high-throughput biology with flexible open-source automation. {\sl Mol Syst Biol}. 2021; 17(3): e9942. \doiUrl{10.15252/msb.20209942}

\ 

Clancy T. {\sl Executive Orders}. New York: Berkley Books, 1997.

\ 

Cohen SN. DNA cloning: A personal view after 40 years. {\sl Proc Nat Acad Sci USA}. 2013; 110(39): 15521-15529. \doiUrl{10.1073/pnas.1313397110}

\ 

Dolan KA, Wang J, Peterson-Withorn C. Forbes World's Billionaires List: the Richest in 2021. 
\url{https://www.forbes.com/billionaires/}

\ 

Dyson FJ. Is Science Mostly Driven by Ideas or by Tools? {\sl Science}. 2012; 338: 1426-1427. \doiUrl{10.1126/science.1232773}

\ 

Ellman, Michael. Soviet repression statistics: some comments. {\sl Europe-Asia Studies}. 2002; 54(7): 1151-1172.

\ 

Gibson DG, et al. Creation of a bacterial cell controlled by a chemically synthesized genome. {\sl Science}. 2010; 329; 52-56.  \doiUrl{10.1126/science.1190719}

\ 

Good IJ. Speculations concerning the first ultraintelligent machine. {\sl Advances in Computers}. 1966; 6: 31-88.
\url{http://www.aeiveos.com/~bradbury/Authors/Computing/Good-IJ/SCtFUM.html}

\ 

Greenberg A. The untold story of NotPetya, the most devastating cyberattack in history. Wired.com. Aug.~22, 2018. \quad https://www.wired.com/story/notpetya-cyberattack-ukraine-russia-code-crashed-the-world/ =  \url{https://bit.ly/3wvQZ5B} 

\ 

Griffin DW. Atmospheric movement of microorganisms in clouds of desert dust and implications for human health. {\sl Clin Microbiol Rev}. 2007; 20: 459-477. \doiUrl{10.1128/CMR.00039-06}

\ 

Hammond AM, Galizi R. Gene drives to fight malaria: curent state and future directions. {\sl Pathogens and Global Health}. 2017; 111: 412-423.

\ 

Herfst S, et al. Airborne transmission of influenza A/H5N1 virus between ferrets. {\sl Science}. 2012; 336:1534-1541. \doiUrl{10.1126/science.1213362}.

\ 

Hobby, Gladys L. {\sl Penicillin: Meeting the Challenge}. New Haven: Yale, 1985. Pages 100-101, 158-159, 234.

\ 

Hoffman DE. {\sl The Dead Hand: the Untold Story of the Cold War Arms Race and its Dangerous Legacy}. New York: Doubleday, 2009.

\ 

Imai M, et al. Experimental adaptation of an influenza H5 HA confers respiratory droplet transmission to a reassortant H5 HA/H1N1 virus in ferrets. {\sl Nature}. 2012; 486: 420-428. \doiUrl{http://dx.doi.org/10.1038/nature10831}

\ 

Jackson RJ, et al. Expression of mouse interleukin-4 by a recombinant ectromelia virus suppresses cytolytic lymphocyte responses and overcomes genetic resistance to mousepox. {\sl Journal of Virology}. 2001; 75: 1205-1210. \doiUrl{10.1128/JVI.75.3.1205-1210.2001}

\ 

Jogalekar A. Chemistry and Biology: Kuhnian or Galisonian? {\sl The Curious Wavefunction} blog. Dec.~20, 2012. \url{https://bit.ly/3BkOoxG} = https://blogs.scientificamerican.com/the-curious-wavefunction/chemistry-galisonian-rather-than-kuhnian/

\ 

Kelly J. {\sl The Great Mortality}. New York: HarperCollins, 2005. Page 281.

\ 

Kim J, Koo B-K, Knoblich JA. Human organoids: model systems for human biology and medicine. {\sl Nature Reviews Molecular Cell Biology}. 2020; 21: 571-584. \doiUrl{10.1038/s41580-020-0259-3}

\ 

Kohn GC. {\sl The Wordsworth Encyclopedia of Plague \&\ Pestilence}. New York: Facts on File, 1995.

\ 

Kupferschmidt K. How Canadian researchers reconstituted an extinct poxvirus for \$100,000 using mail-order DNA. Sciencemag.org. July 6, 2017.  \doiUrl{10.1126/science.aan7069}

\ 

Leventhal DS, et al. Immunotherapy with engineered bacteria by targeting the STING pathway for anti-tumor immunity. {\sl Nature Communications}. 2020; 11: 2739.\\ \doiUrl{10.1038/s41467-020-16602-0}

\ 

Liu H-Y, Lauta KC, Maas MM. Governing boring apocalypses: a new typology of existential vulnerabilities and exposures for existential risk research. {\sl Futures}. 2018; 102: 6-19. \doiUrl{10.1016/j.futures.2018.04.009}

\ 

Monie DD, et al. Synthetic and systems biology principles in the design of programmable oncolytic virus immunotherapies for glioblastoma. {\sl Neurosurg Focus}. 2021; 50(2): E10. \doiUrl{10.3171/2020.12.FOCUS20855}

\ 

Naderi M, et al. Binding site matching in rational drug design: algorithms and applications. {\sl Brief Bioinform}. 2019 Nov 27;20(6):2167-2184. \doiUrl{10.1093/bib/bby078} 

\ 

Nakashima E, Torbati Y, Englund W. Ransomware attack leads to shutdown of major U.S. pipeline system. Washington Post. May~8, 2021. \url{https://www.washingtonpost.com/business/2021/05/08/cyber-attack-colonial-pipeline/}

\ 

Nassar, Sylvia. {\sl A Beautiful Mind:\ The Life of Mathematical Genius and Nobel Laureate John Nash}. New York: Simon \&\ Schuster Touchstone, 1998.

\ 

Noyce RS, Evans DH. Synthetic horsepox viruses and the continuing debate about dual use research. {\sl PLoS Pathog}. 2018; 14: e1007025. \doiUrl{10.1371/journal.ppat.1007025}

\ 

Oldstone MBA. {\sl Viruses, Plagues, \&\ History}. Oxford: Oxford, 2010. Pages 61-63.

\ 

Palmer AW. On the Trail of the Robocall King. {\sl Wired}. March 25, 2019. \url{https://www.wired.com/story/on-the-trail-of-the-robocall-king/}

\ 

Parks, Jerry M.; Smith, Jeremy C. ``How to discover antiviral drugs quickly.'' {\sl N Engl J Med}. 2020; 382:2261-2264. \doiUrl{10.1056/NEJMcibr2007042}

\ 

Plum, Fred; Posner, Jerome B. {\sl The Diagnosis of Stupor and Coma}. 3rd ed. Philadelphia: F.A.\ Davis, 1980. Page 9.

\ 

Putman P. Script kiddie: unskilled amateur or dangerous hackers? Accessed June 27, 2021. \url{https://www.uscybersecurity.net/script-kiddie/}

\ 

Sagan, Carl (ed.). {\sl Communication with Extraterrestrial Intelligence}. Cambridge, MA: MIT Press, 1973.

\ 

Sandberg A, Bostrom N. ``Global Catastrophic Risks Survey.'' Technical Report \#2008-1. Future of Humanity Institute, Oxford University. 2008. (Accessed June 27, 2021.) \url{https://www.fhi.ox.ac.uk/reports/2008-1.pdf}

\ 

Schlosser E. {\sl Command and Control: Nuclear Weapons, the Damascus Accident, and the Illusion of Safety}. New York: Penguin, 2013.

\ 

Schr\"odinger E. {\sl Nature and the Greeks -and- Science and Humanism}. Cambridge, UK: Cambridge University Press, 1996.

\ 

Schutgens F, Clevers H. Human Organoids: Tools for Understanding Biology and Treating Diseases. {\sl Annual Review of Pathology: Mechanisms of Disease}. 2020; 15(1): 211-234. \doiUrl{10.1146/annurev-pathmechdis-012419-032611}

\ 

Sheikh A, et al. SARS-CoV-2 Delta VOC in Scotland: demographics, risk of hospitalization, and vaccine effectiveness. {\sl The Lancet}. 2021: 397; 2461-2462

\ 

Shirley JM, de Jong YP, Terhorst C, Herzog RW. Immune Responses to Viral Gene Therapy Vectors. {\sl Molecular Therapy}. 2020; 28(3): 709-722. \doiUrl{10.1016/j.ymthe.2020.01.001}

\ 

Sieburg HB, Baray C, Kunzelman KS. Testing HIV molecular biology in {\sl in silico} physiologies. Proc Int Conf Intell Syst Mol Biol. 1993;1:354-361.

\ 

Sotos JG. Biotechnology and the lifetime of technical civilizations. {\sl International Journal of Astrobiology}. 18 (5): 445-454. \doiUrl{10.1017/S1473550418000447}

\ 

Stewart, George R. {\sl Earth Abides}. New York: Random House, 1949.

\ 

Strickland E. The robot revolution comes to synthetic biology. {\sl IEEE Spectrum}. 2016; 53(12): 9-11. \url{https://ieeexplore.ieee.org/stamp/stamp.jsp?arnumber=7761863}

\ 

Tainter, Joseph A. {\sl The Collapse of Complex Societies}. Cambridge, UK: Cambridge University Press, 1988. Page 4.

\ 

Thacker EE, Timares L, Matthews QL. Strategies to overcome host immunity to adenovirus vectors in vaccine development. Expert Rev Vaccines. 2009 Jun; 8(6): 761-777. \doiUrl{10.1586/erv.09.29} 

\ 

Torres P.  Agential risks: a comprehensive introduction. {\sl Journal of Evolution and Technology}. 2016; 26(2): 31-47. \url{https://jetpress.org/v26.2/torres.htm}

\ 

van Melle W, Shortliffe EH, Buchanan BG. EMYCIN: A knowledge engineer's tool for constructing rule-based expert systems. Chapter 15 (pages 302-313) in: {\sl Rule-Based Expert Systems: The MYCIN Experiments of the Stanford Heuristic Programming Project}. Reading, MA: Addison-Wesley, 1984. \url{https://www.aaai.org/Papers/Buchanan/Buchanan17.pdf}

\ 

Wade N. The origin of COVID: Did people or nature open Pandora's box at Wuhan? {\sl Bulletin of the Atomic Scientists} online. May~5, 2021. \url{https://thebulletin.org/2021/05/the-origin-of-covid-did-people-or-nature-open-pandoras-box-at-wuhan/}

\ 

Wang Y, et al. In silico ADME/T modelling for rational drug design. {\sl Q Rev Biophys}. 2015 Nov;48(4):488-515. \doiUrl{10.1017/S0033583515000190}

\ 

Warrell DA. Snake bite. {\sl Lancet}. 2010; 375: 77-88. \doiUrl{10.1016/S0140-6736(09)61754-2}

\ 

Wired Staff. WIRED25: Stories of People Who Are Racing to Save Us: Emily Leproust. {\sl Wired Magazine}. November 2019. \\ \url{https://www.wired.com/story/wired25-stories-people-racing-to-save-us/}

\end{document}